\documentclass[twocolumn,aps,prl,superscriptaddress,showpacs,floatfix,longbibliography]{revtex4-1}
\usepackage{url}
\usepackage{cancel}
\usepackage[colorlinks,linkcolor=blue,citecolor=blue,filecolor=black,urlcolor=blue]{hyperref}
\usepackage{epsfig,graphics}
\usepackage{graphicx}
\usepackage{dcolumn}
\usepackage{bm}
\usepackage[usenames]{color}
\usepackage{amssymb}
\usepackage{amsmath}
\usepackage{multirow}
\usepackage{float}
\usepackage{harpoon}
\usepackage{MnSymbol}
\usepackage{appendix}
\usepackage{color}
\usepackage{hyperref}
\usepackage{cleveref}

\newcommand{\sqrtsnn}{\mbox{$\sqrt{s^{}_{\mathrm{NN}}}$}}

\newcommand{\etaref} {\eta_{\mathrm{ref}}}
\newcommand{\lr}[1]{\left\langle #1\right\rangle}
\newcommand{\llrr}[1]{\left\llangle #1\right\rrangle}

\newcommand\be{\begin{equation}}
\newcommand\ee{\end{equation}}

\begin{document}

\title{New constraints on equation of state of hot QCD matter}

\author{\small Lu-Meng Liu}\email[Correspond to\ ]{liulumeng@fudan.edu.cn}
\affiliation{Physics Department and Center for Particle Physics and Field Theory, Fudan University, Shanghai 200438, China}

\author{\small Jinhui Chen}\email[Correspond to\ ]{chenjinhui@fudan.edu.cn}
\affiliation{Key Laboratory of Nuclear Physics and Ion-beam Application (MOE), Fudan University, Shanghai 200433, China}
\affiliation{Shanghai Research Center for Theoretical Nuclear Physics, National Natural Science Foundation of China and Fudan University, Shanghai 200438, China}

\author{\small Xu-Guang Huang}\email[Correspond to\ ]{huangxuguang@fudan.edu.cn}
\affiliation{Physics Department and Center for Particle Physics and Field Theory, Fudan University, Shanghai 200438, China}
\affiliation{Key Laboratory of Nuclear Physics and Ion-beam Application (MOE), Fudan University, Shanghai 200433, China}
\affiliation{Shanghai Research Center for Theoretical Nuclear Physics, National Natural Science Foundation of China and Fudan University, Shanghai 200438, China}

\author{\small Jiangyong Jia}\email[Correspond to\ ]{jiangyong.jia@stonybrook.edu}
\affiliation{Department of Chemistry, Stony Brook University, Stony Brook, NY 11794, USA}
\affiliation{Physics Department, Brookhaven National Laboratory, Upton, NY 11976, USA}

\author{\small Chun Shen}\email[Correspond to\ ]{chunshen@wayne.edu}
\affiliation{Department of Physics and Astronomy, Wayne State University, Detroit, Michigan 48201, USA}

\author{\small Chunjian Zhang}\email[Correspond to\ ]{chunjianzhang@fudan.edu.cn}
\affiliation{Key Laboratory of Nuclear Physics and Ion-beam Application (MOE), Fudan University, Shanghai 200433, China}
\affiliation{Shanghai Research Center for Theoretical Nuclear Physics, National Natural Science Foundation of China and Fudan University, Shanghai 200438, China}

\begin{abstract}
The longitudinal structure of the quark-gluon plasma (QGP) remains a key challenge in heavy-ion physics. In this Letter, we propose a novel observable, event-by-event mean transverse momentum fluctuations $\mathrm{Var}_{\langle p_T \rangle}$, which is sensitive to the local pressure gradients and serves as a probe of longitudinal dynamics in the initial state of QGP. We demonstrate that the covariance of averaged transverse momentum at two rapidities $\mathrm{Cov}_{\langle p_T \rangle}(\eta_1, \eta_2)$ and its associated decorrelation measures, $R_{p_T}(\eta_1, \eta_2)$ and $r_{p_T}(\eta, \eta_\mathrm{ref})$, exhibit strong sensitivity to the stiffness of equation of state (EoS) of QGP, while showing negligible dependence on the QGP transport coefficients. This distinctive behavior, revealed through state-of-the-art (3+1)-dimensional hydrodynamic simulations, establishes a powerful approach for constraining the EoS of QCD matter. In the meantime, our results provide new insights into the longitudinal structure of the QGP and its properties under high baryon density.
\end{abstract}

\maketitle

\textbf{Introduction.} Determining the equation of state (EoS) in quantum chromodynamics (QCD) matter at finite densities remains a central challenge in nuclear physics, with profound implications spanning from compact stars to heavy-ion collisions~\cite{Danielewicz:2002pu,Li:2008gp,Sorensen:2023zkk}. The EoS stiffness manifests distinctly across physical systems: in astrophysics, it governs neutron star structure through the Tolman-Oppenheimer-Volkoff equations~\cite{Lattimer:2012nd,Oertel:2016bki}; while in heavy-ion collisions, the EoS directly controls the quark-gluon plasma (QGP) fireball expansion dynamics, with stiffer equations of state (characterized by larger sound velocity $c_s^2$) generating larger acceleration from pressure gradients that drive faster hydrodynamic expansion, ultimately manifesting in enhanced anisotropic flow in the final-state particle distributions.

In relativistic heavy-ion collisions, performed at the Relativistic Heavy Ion Collider (RHIC) and the Large Hadron Collider (LHC), the creation of QGP provides a unique opportunity to study the fundamental properties of QCD under extreme conditions~\cite{Shuryak:1980tp,ALICE:2004fvi,STAR:2005gfr,PHENIX:2004vcz,Gardim:2019xjs}. Understanding the evolution of QGP from initial state to final-state momentum-space anisotropic flow is a central theme~\cite{Ollitrault:1992bk,Sorge:1998mk,Bilandzic:2010jr,Heinz:2013th}. This evolution is driven by pressure gradient-driven hydrodynamic expansion. For instance, the event-averaged mean transverse momentum \( \llrr{p_{T}} \), driven by radial flow, correlates positively with the inverse transverse size of the initial energy density profile~\cite{Bozek:2012fw,Giacalone:2020dln,Cao:2021zhy,Schenke:2020uqq,Jia:2022ozr,STAR:2024wgy,STAR:2025rot}. Studies employing various EoS at RHIC and LHC energies indicate that a larger speed of sound generally leads to increased anisotropic flow and a higher mean transverse momentum at mid-rapidity~\cite{Sangaline:2015isa,Nijs:2023bzv,Gong:2024lhq}. Recently, there have been proposals to measure the speed of sound through the relation of $c_s^2 = d P/d e\sim d\,\mathrm{ln} \llrr{p_{T}} / d\,\mathrm{ln} N_{ch}$ with collisions at a fixed volume in ultra-central centrality~\cite{Gardim:2019brr,CMS:2024sgx}.
While transverse dynamics have been extensively studied, longitudinal dynamics also play a critical role in shaping QGP’s evolution, particularly in low-energy regimes such as the Beam Energy Scan (BES) program where the Bjorken scaling hypothesis (\( \eta_s \approx y \)) breaks down with $\eta_s = \frac{1}{2} \ln \left( \frac{t+z}{t-z}\right)$ and $y = \frac{1}{2} \ln \left( \frac{E+p_z}{E-p_z}\right)$ representing space-time and flow rapidity, respectively. Furthermore, event-by-event (EbE) fluctuations in the initial state are also expected to induce rapidity-dependent inhomogeneities in entropy deposition, which may decorrelate mean \( \llrr{p_{T}} \) at different rapidity bins~\cite{Bozek:2010vz,Bozek:2015bna,Jia:2014ysa}.

Relativistic viscous hydrodynamics simulations successfully describe experimental observables in heavy-ion collisions, with their predictive power fundamentally relying on the EoS that relates local energy density to pressure. At LHC and the top RHIC energies, where the net baryon density and chemical potential $\mu_{B}$ approach zero, the EoS can be calculated from the first principles using lattice QCD techniques~\cite{Borsanyi:2010cj,Bazavov:2009zn,HotQCD:2014kol,HotQCD:2018pds}. As collision energies decrease and baryon stopping becomes significant, the finite $\mu_{B}$ regime presents both challenges and opportunities~\cite{Shen:2025unr,Chen:2025eeb}. This region allows exploration of QGP properties at lower temperatures and higher baryon densities, but lies beyond reliable lattice QCD calculations due to the sign problem~\cite{Karsch:2001cy,Muroya:2003qs,Bedaque:2017epw,Ratti:2018ksb}. Current understanding of this crucial region mostly relies on effective QCD models~\cite{Fukushima:2013rx, Liu:2021gsi,Fu:2019hdw,Gao:2020qsj}, multimessenger astronomy~\cite{Tsang:2023vhh}, and Taylor expansion based on higher-order susceptibilities at vanishing net baryon chemical potential~\cite{HotQCD:2018pds,Monnai:2019hkn}. Despite these efforts, substantial uncertainties remain in the finite-density EoS, underscoring the importance of heavy-ion collisions for mapping the complete QCD phase diagram under extreme conditions~\cite{Fukushima:2010bq,Luo:2017faz,Bzdak:2019pkr,Chen:2024aom}.

Prior theoretical and experimental studies have explored longitudinal decorrelations in high-energy nuclear collisions, including the pseudorapidity dependence of anisotropic flow~\cite{Petersen:2011fp,Pang:2012he,Jia:2017kdq,Pang:2014pxa,Zhu:2024tns,Wu:2021hkv,Zhang:2024bcb,Jia:2024xvl,CMS:2015xmx,ATLAS:2020sgl,Nie:2019bgd}, forward-backward decorrelations of transverse momentum and their connection to flow~\cite{Bozek:2016yoj,Chatterjee:2017mhc}, as well as multiplicity decorrelations~\cite{ATLAS:2016rbh,Jia:2015jga,Jia:2020tvb}. These observations provide crucial first insights into the longitudinal structure of the QGP. Yet, the intrinsic longitudinal dynamics of the QGP and how they are influenced by the equation of state remain largely unexplored. Recent studies indicate that the transverse momentum longitudinal correlation $G_2(\eta_1, \eta_2)$ originates from hydrodynamic response to pre-equilibrium correlations~\cite{STAR:2011iio,ALICE:2019smr,Savchuk:2024ycj}.

In this Letter, we employ a full (3+1)D viscous hybrid hydrodynamic framework to simulate \(^{197}\)Au+\(^{197}\)Au collisions at $\sqrt{s_{\rm NN}}=$ 19.6 GeV, the BES program energy where both EoS at such finite baryon density effects and non-Bjorken longitudinal expansion are significant. We propose novel measures of \( \langle p_{\rm T} \rangle \) decorrelations, the Pearson coefficient \( R_{p_{\rm T}}(\eta_1, \eta_2) \) and the three-bin decorrelation \( r_{p_{\rm T}}(\eta, \eta_{\rm ref})\). These observables is highly sensitive to local smearing of pressure, in contrast to the decorrelation of anisotropic flow in large-scale structures. This sensitivity allows for the systematic quantification of longitudinal fluctuation patterns and provides new constraints on EoS of QCD matter.

\textbf{Observables.} The \(p_T\) decorrelation that origin from the EbE fluctuation of the size and entropy of the QGP initial state along the pseudorapidity direction can be characterized by~\cite{Chatterjee:2017mhc}
\begin{eqnarray}\label{eq:RpT}
R_{p_T}\left({\eta_1},{\eta_2}\right)=\frac{\operatorname{Cov}_{\lr{p_T}} \left({\eta_1},{\eta_2}\right)}{\sqrt{\operatorname{Var}_{\lr{p_T}}\left({\eta_1}\right)}\sqrt{\operatorname{Var}_{\lr{p_T}}\left({\eta_2}\right)}},
\end{eqnarray}
where 
\begin{eqnarray}\label{eq:covpT}
\operatorname{Cov}_{\lr{p_T}} \left({\eta_1},{\eta_2}\right) 
&=& \left\langle\left(\lr{p_T}_{\eta_1}-\llrr{p_T}_{\eta_1}\right)\left(\lr{p_T}_{\eta_2}-\llrr{p_T}_{\eta_2}\right)\right\rangle, \nonumber\\
\operatorname{Var}_{\lr{p_T}}\left({\eta}\right) &=& \operatorname{Cov}_{\lr{p_T}} \left({\eta},{\eta}\right) ,
\end{eqnarray}
are the covariance and variance of $\lr{p_T}_{\eta}$, respectively.
$R_{p_T}\left({\eta_1},{\eta_2}\right)$ characterizes the  correlations of $\lr{p_T}$ between two pseudorapidity bins $\eta_1$ and $\eta_2$, averaged over the ensemble of events, satisfying $R_{p_T}\left({\eta},{\eta}\right)=1$. 
To suppress nonflow contributions, we employ a three-bin correlator, analogous to the method used in anisotropic flow analysis~\cite{CMS:2015xmx}, defined as:
\begin{eqnarray}\label{eq:rpT}
r_{p_T}\left({\eta},{\etaref}\right) = \frac{\operatorname{Cov}_{\lr{p_T}} \left({-\eta},{\etaref}\right)}{\operatorname{Cov}_{\lr{p_T}} \left({\eta},{\etaref}\right)} = \frac{R_{p_T}\left({-\eta},{\etaref}\right)}{R_{p_T}\left({\eta},{\etaref}\right)}.
\end{eqnarray}
The second equality is a direct consequence of the variance symmetry $\operatorname{Var}_{\lr{p_T}}\left({\eta}\right) = \operatorname{Var}_{\lr{p_T}}\left({-\eta}\right)$ in symmetric collisions.
In the above, the EbE mean transverse momentum $\lr{p_T}\left(\eta\right)$ and its event average $\llrr{p_T}\left(\eta\right)$ are given by
\begin{eqnarray}
\lr{p_T}_\eta &=& \frac{1}{N_\eta}\sum_{i=1}^{N_\eta} p_{\mathrm{T},i}, ~~~~\llrr{p_T}_\eta = \frac{1}{N_{ev}}\sum_{j=1}^{N_{ev}} \lr{p_T}_{\eta,j} \nonumber
\end{eqnarray}
with $N_\eta$ counts charged particles in the $\eta$ bin per event and $N_{ev}$ is the total number of events.

\begin{figure}[ht]
\centering
\vspace*{-0.3cm}
\includegraphics[width=0.75\linewidth]{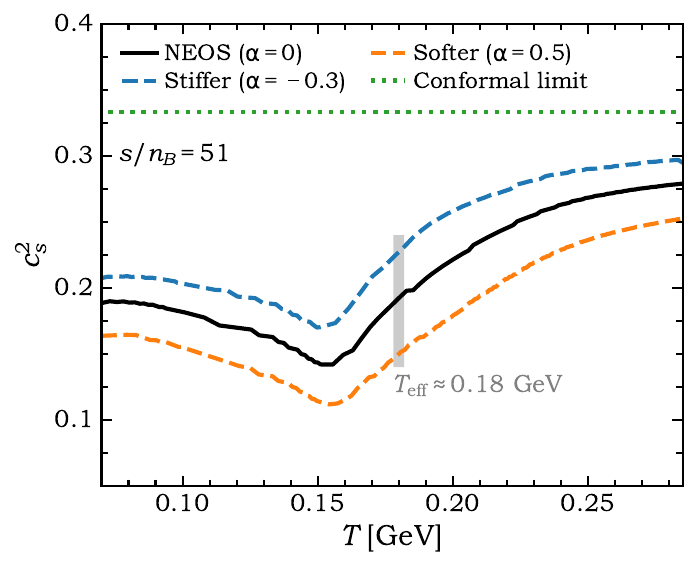}
\vspace*{-0.4cm}
\caption{Squared speed of sound ($c_s^2$) as a function of temperature, computed under the isentropic condition \( s/n_B = 51 \) for $^{197}$Au+$^{197}$Au collisions at \( \sqrt{s_{\rm NN}} = 19.6\ \mathrm{GeV} \) using different EoS parametrizations. The fireball effective temperature can be estimated via the approximate relation $ T_\mathrm{eff} \approx \llrr{p_T} / 3 \approx 0.18~\mathrm{GeV}$~\cite{Gardim:2019xjs}.}
\label{fig:cs2}
\end{figure}

\textbf{Equation of state in QCD matter.} To systematically investigate the effects of the equation of state, we introduce a parametric modification of the pressure \( P_{\mathrm{NEOS}}(T, \mu_B) \) provided by lattice-QCD-based NEOS-BQS equation of state at finite baryon densities~\cite{Monnai:2019hkn}. This NEOS-BQS equation of state combines the pressure from lattice QCD calculations~\cite{HotQCD:2014kol} with that of the hadron resonance gas model. The finite-density behavior is systematically incorporated via a Taylor expansion approach, utilizing higher-order baryon number susceptibilities calculated at zero net baryon chemical potential from state-of-the-art lattice QCD results~\cite{HotQCD:2018pds}. A free parameter $\alpha(\ge -1)$ is introduced to control the stiffness of the EoS via the transformation
\begin{equation}\label{eq:Palpha}
P_\alpha(T, \mu_B) = P_{\mathrm{NEOS}}(T, \mu_B) \left[ \frac{P_{\mathrm{NEOS}}(T, \mu_B)}{T^4} \right]^\alpha.
\end{equation}
This form maintains thermodynamic consistency while allowing for systematic variation of the EoS stiffness, and simultaneously prevents the speed of sound from exceeding the conformal limit, as shown in Fig.~\ref{fig:cs2}. The corresponding entropy density, conserved charge densities, and energy density are derived as
\begin{align}
s_\alpha &= \frac{\partial P_\alpha}{\partial T}, ~~~~~~~n_{J,\alpha} = \frac{\partial P_\alpha}{\partial \mu_J}, \\
e_\alpha &= -P_\alpha + T s_\alpha + \sum_J \mu_J n_{J,\alpha},
\end{align}
where $J$ denotes the conserved charges. Detailed derivations can be found in the appendix.

Figure~\ref{fig:cs2} shows the speed of sound squared along the constant entropy-per-baryon line \( s/n_B = 51 \), a condition relevant for the BES program and corresponding to $^{197}$Au+$^{197}$Au collisions at \( \sqrt{s_{\rm NN}} = 19.6\ \mathrm{GeV} \)~\cite{Guenther:2017hnx}.
A stiffer equation of state with \( \alpha = -0.3 \) yields \( c_s^2(T = 0.18~\mathrm{GeV}, s/n_B = 51) = 0.225 \), which is about 15\% larger than the baseline NEOS-BQS value of 0.19 at the same temperature and entropy-per-baryon. In contrast, a softer EoS with \( \alpha = 0.5 \) gives a lower value of 0.15. By varying the parameter \( \alpha \), this parametrization generates a family of equations of state with controlled stiffness, enabling systematic investigations of EoS effects in hydrodynamic simulations.

\textbf{Hydrodynamic simulations.} 
We employ a (3+1)D dynamical initialization model (3D-Glauber)~\cite{Shen:2017bsr,Shen:2022oyg} coupled with the hybrid framework of relativistic viscous hydrodynamics (MUSIC)~\cite{Schenke:2010nt,Schenke:2010rr,Paquet:2015lta} + hadronic transport (UrQMD)~\cite{Bass:1998ca,Bleicher:1999xi} to simulate the state-of-the-art (3+1)D heavy-ion collisions. 
In this framework, the hydrodynamic equation of motion with source terms~\cite{Shen:2017bsr,Shen:2022oyg} 
\begin{eqnarray}
   \partial_\mu T^{\mu\nu} &=& J^\nu_{\text{source}}, \\
   \partial_\mu J^\mu &=& \rho_{\text{source}},
\end{eqnarray}
are solved with the EoS input shown in Fig.~\ref{fig:cs2}. 

In our viscous hydrodynamics framework, the specific shear and bulk viscosities are normalized by $T/(e+P)$ rather than by the local entropy density $s$. This normalization scheme provides enhanced numerical stability for handling the viscous relaxation times $\tau_\pi$ and $\tau_\Pi$. The normalized viscosities, $\eta T / (e + P)$ and $\zeta T / (e + P)$, converge to the conventional definitions of $\eta/s$ and $\zeta/s$ in the limit of vanishing net baryon density. We explore several constant values for the specific shear viscosity, $\eta T / (e + P) = \{0.08, \mathbf{0.15}, 0.22\}$, while the specific bulk viscosity $\zeta T / (e + P)$ follows the parameterization~\cite{Schenke:2020mbo}
\begin{equation}
\frac{\zeta T }{e + P} = 
\begin{cases} 
B_\mathrm{norm} \exp \left[ -\dfrac{(T - T_\mathrm{peak})^2}{B_1^2} \right], & T < T_\mathrm{peak}, \\
B_\mathrm{norm} \exp \left[ -\dfrac{(T - T_\mathrm{peak})^2}{B_2^2} \right], & T > T_\mathrm{peak},
\end{cases}
\end{equation}
with parameter values $B_\mathrm{norm} = \{0.1, \mathbf{0.13}, 0.18\}$, $B_1 = 0.01$ GeV, $B_2 = 0.12$ GeV, and $T_\mathrm{peak} = 0.16$ GeV.

To examine the impact of different equations of state and transport coefficients, we systematically explore a range of viscosity parameters for each EoS, with specific combinations detailed in Table~\ref{tab:parameters}.  
Additional model parameters are adopted from the maximum-likelihood parameter set in Ref.~\cite{Jahan:2024wpj}. 
These parameter variations are calibrated to reproduce the pseudorapidity ($\eta$) dependence of anisotropic flow observables.
The rapidity distribution of charged hadron yields is anticipated to exhibit relative insensitivity to the EoS variations, as the dominant contribution to the system entropy originates from pre-hydrodynamic stages.
\begin{table}[t]
    \centering
    \caption{Parameter sets for the equation of state and viscosities used in the simulations. The squared speed of sound $c_s^2$ is parametrized as $c_s^2(T=0.18~\mathrm{GeV}, s/n_B=51)$. Columns 4 and 5 correspond to the shear and bulk viscosity parameters, respectively.}
    \label{tab:parameters}
	\renewcommand{\arraystretch}{1.3}
	\setlength{\tabcolsep}{1.4mm}
    \begin{tabular}{ccccc}
        \hline\hline
        & EoS: $P_\alpha(T,\mu_B)$ & $c_s^2$ & $\eta T / (e + P)$ & $B_\text{norm}$ \\ 
        \hline
        NEOS & NEOS-BQS $(\alpha=0)$ & 0.19 &$\mathbf{0.15}$ & $\mathbf{0.13}$ \\ 
        Stiff 1 & Stiffer EoS $(\alpha=-0.3)$ & 0.225 &$\mathbf{0.15}$ & $\mathbf{0.13}$ \\ 
        Stiff 2 & Stiffer EoS $(\alpha=-0.3)$ & 0.225 &0.22 & 0.18 \\ 
        Soft 1 & Softer EoS $(\alpha=0.5)$ & 0.15 &$\mathbf{0.15}$ & $\mathbf{0.13}$ \\ 
        Soft 2 & Softer EoS $(\alpha=0.5)$ & 0.15 &0.08 & 0.10 \\ 
        \hline\hline
    \end{tabular}
\end{table}

\begin{figure}[ht]
\centering
\vspace*{-0.2cm}
\includegraphics[width=1.\linewidth]{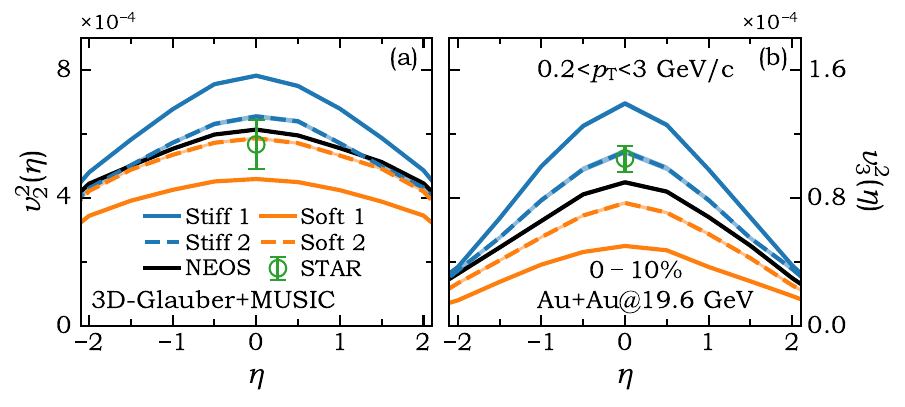}
\vspace*{-0.9cm}
\caption{The pseudorapidity \(\eta\) distribution of anisotropic flow coefficients $v_2^2$ (a) and $v_3^2$ (b) on  in 0-10\% central $^{197}$Au+$^{197}$Au collisions at 19.6 GeV with different EoS and viscosity parameters presented in Table~\ref{tab:parameters}. The STAR data is taken from Refs.~\cite{STAR:2017idk,STAR:2016vqt}.}
\label{fig:flow}
\end{figure}

\textbf{Results.} Figure~\ref{fig:flow} presents the pseudorapidity dependence of squared anisotropic flow coefficients \(v_2^2\) and \(v_3^2\) for 0-10\% central $^{197}$Au+$^{197}$Au collisions at \(\sqrtsnn=19.6\) GeV, with charged particles selected in the transverse momentum range \(0.2<p_{\rm{T}}<3\) GeV/\(c\). The Stiff 1 case exhibits systematically larger flow coefficients $v_n^2$ across the full \(\eta\) range compared to the NEOS results. This amplification is consistent with hydrodynamic evolution associated with the increased $c_s$. Conversely, the Soft 1 scenario leads to smaller flow magnitudes due to slower expansion. We also present the Stiff 2 and Soft 2 cases where the modified shear and bulk viscosities are adjusted to compensate for the EoS changes, effectively reproducing the anisotropic flow patterns of the NEOS calculation. These comparisons reveal the competing effects between the EoS and viscous transport coefficients during QGP evolution. Our model with the NEOS EoS provides a reasonable description of the STAR $v_n^2$ data within the experimental uncertainties~\cite{STAR:2017idk,STAR:2016vqt}, establishing a baseline for our subsequent investigations of the EoS sensitivity.

\begin{figure}[ht]
\centering
\vspace*{-0.2cm}
\includegraphics[width=1.\linewidth]{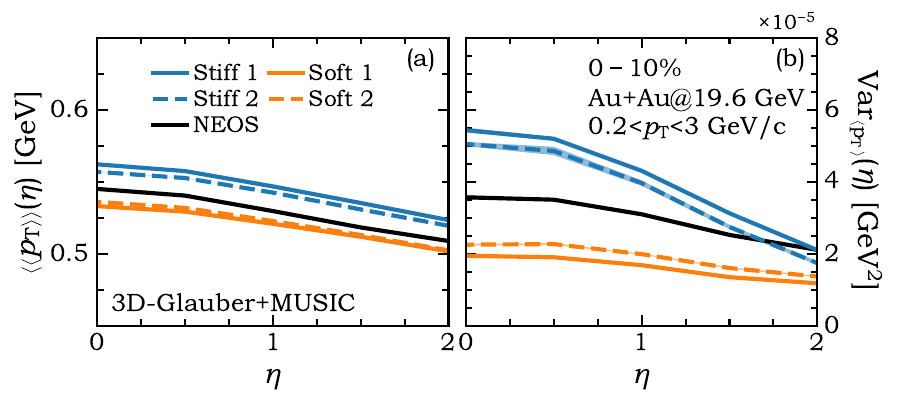}
\vspace*{-0.8cm}
\caption{The pseudorapidity \(\eta\) distribution of mean transverse momentum $\llrr{p_T}_\eta$ (a) and it's covariance \(\operatorname{Var}_{\lr{p_T}} \left({\eta}\right)\) (b) in 0-10\% central $^{197}$Au+$^{197}$Au collisions at 19.6 GeV.}\label{fig:varpT}
\end{figure}

Figure~\ref{fig:varpT} illustrates the \(\eta\) distribution of $\left\llangle p_T\right\rrangle_\eta$ (panel a) and corresponding variance $\text{Var}_{\langle p_T\rangle}(\eta)$ (panel (b)) as defined in Eq.~(\ref{eq:covpT}). We observe that the stiffer equation of state (Stiff~1) yields enhanced $\left\llangle p_T\right\rrangle_\eta$ by approximately 0.02~GeV, whereas the softer EoS (Soft~1) exhibits the opposite trend. Furthermore, larger bulk viscosity generally suppresses $\left\llangle p_T\right\rrangle_\eta$, while smaller viscosity enhances it.
Figure~\ref{fig:varpT}(b) shows that $\text{Var}_{\langle p_T\rangle}(\eta)$ depends mainly on the equation of state and less on viscosity, particularly at mid-rapidity. Since viscous effects represent higher-order corrections to the hydrodynamic evolution, this observable provides a cleaner probe of the EoS. The stiffer EoS (Stiff~1/2) produces larger $p_T$ fluctuation, while the softer EoS (Soft~1/2) yields smaller $p_T$ fluctuations. These results are consistent with previous $p_T$ fluctuations and the speed of sound, where $\sqrt{\operatorname{Var}_{\lr{p_T}}} / \left\llangle p_T\right\rrangle \sim c_s^2 $~\cite{Giacalone:2020lbm}.

\begin{figure}[ht]
\centering
\vspace*{-0.2cm}
\includegraphics[width=1.0\linewidth]{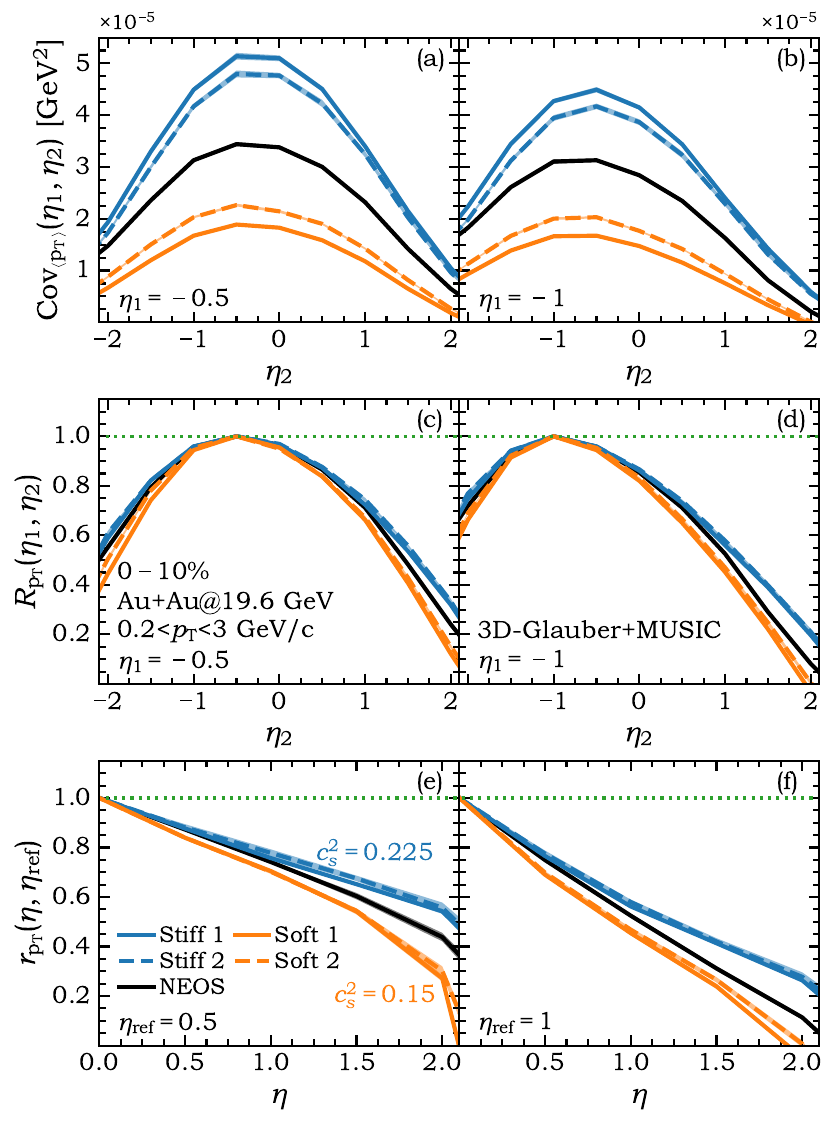}
\vspace*{-0.5cm}
\caption{Mean transverse momentum decorrelation coefficients in 0--10\% central $^{197}$Au+$^{197}$Au collisions. Panels (a) and (b) show $\operatorname{Cov}_{\lr{p_T}}(\eta_1, \eta_2)$ with $\eta_1 = -0.5$ and $\eta_1 = -1$, respectively; Panels (c) and (d) show $R_{p_T}(\eta_1, \eta_2)$ with $\eta_1 = -0.5$ and $\eta_1 = -1$, respectively; Panels (e) and (f) present $r_{p_T}(\eta, \etaref)$ with $\etaref = 0.5$ and $\etaref = 1$, respectively.}\label{fig:capRpT}
\end{figure}

Figure~\ref{fig:capRpT}(a,b) displays the $\text{Cov}_{\langle p_T\rangle}(\eta_1, \eta_2)$, similar to $\text{Var}_{\langle p_T\rangle}(\eta)$ in Figure~\ref{fig:varpT}(b), depends mainly on the equation of state and less on viscosity, particularly at large pseudorapidities ($|\eta| \sim 1$) where almost independent on the viscosity. The enhanced EoS sensitivity at forward rapidities originates from longitudinal dynamics during expansion, particularly at lower collision energies where significant violations of boost invariance occur and substantial longitudinal entropy gradients exist in the initial state. Therefore the $p_T$ covariance provides a cleaner probe of the EoS at large pseudorapidities compared to midrapidity measurements, where viscosity plays a more prominent role.

Figure~\ref{fig:capRpT}(c,d) displays the pseudorapidity decorrelation coefficient $R_{p_{\rm{T}}}(\eta_1,\eta_2)$ [Eq.~(\ref{eq:RpT})] for 0--10\% central $^{197}$Au+$^{197}$Au collisions at 19.6~GeV across different $\eta_1$ values. The results reveal systematic enhancement of decorrelation with increasing pseudorapidity gap $\Delta\eta = |\eta_1-\eta_2|$, showing approximately 50\% deviation from unity at $\Delta\eta \sim 2$. This indicates substantial boost invariance violation that becomes particularly pronounced at lower collision energies and large rapidity intervals.
Different with the $p_T$ covariance in Fig.~\ref{fig:capRpT}(a,b), the decorrelation exhibits clear EoS dependence with large $\eta$ gap, with approximately 30\% variation between different equations of state at $\eta_1 = -1$ and $\eta_2 = 1.5$ [Fig.~\ref{fig:capRpT}(d)]. A stiffer EoS leads to weaker decorrelation, as it produces a more sharply localized distribution of pressure gradients that exhibit less rapidity-dependent de-correlation. In contrast, a softer EoS results in stronger decorrelation. Transport coefficients play a subdominant role, contributing only about 5--10\% variation, consistent with previous analyses~\cite{Savchuk:2024ycj}. 
This systematic dependence is consistently observed across different reference pseudorapidity bins [Fig.~\ref{fig:capRpT}(c) at $\eta_1 = -0.5$] and further confirmed by the three-bin decorrelation coefficient $r_{p_{\rm{T}}}(\eta,\eta_{\rm{ref}})$ [Eq.~(\ref{eq:rpT})] shown in Fig.~\ref{fig:capRpT}(e,f), where the effects appear even more pronounced. 
These comprehensive results not only underscore the crucial role of longitudinal dynamics in heavy-ion collisions but also establish forward rapidity decorrelation as a powerful new tool for probing the QCD equation of state under extreme conditions. We have verified that longitudinal flow decorrelations in large-scale structure exhibit minimal dependence on both the EoS and viscosity (See details in appendix), highlighting the unique sensitivity of $p_T$ decorrelation to the equation of state~\cite{Nie:2019bgd}. The above findings demonstrate that longitudinal decorrelation with large pseudorapidity gaps provides a sensitive probe of the equation of state at finite baryon densities. 

\textbf{Summary.}\label{sec:summary} In summary, we have investigated transverse momentum decorrelation in pseudorapidity for 0-10\% central \(^{197}\)Au+\(^{197}\)Au collisions at \sqrtsnn = 19.6 GeV using a (3+1)D hydrodynamic framework. Our results show significant longitudinal decorrelation effect (\(R_{p_{\rm T}}\) and \(r_{p_{\rm T}}\)) with approximately 50\% deviation from unity across \(\Delta\eta\sim 2\), challenging traditional boost-invariance assumptions and highlighting the nature of 3D QGP initial conditions. The decorrelation exhibits pronounced dependence on the stiffness of the EoS, where a stiffer EoS results in weaker decorrelation and a softer EoS leads to stronger decorrelation, while demonstrating negligible sensitivity to viscosity parameters. Our results reveal the particular importance of finite baryon density effects and longitudinal dynamics in low energy. These findings establish  \( \langle p_{\rm T} \rangle \) decorrelations as a powerful probe of QGP initial conditions, providing new constraints on the EoS of QCD matter. Current high-statistical data of BES-II collision energies~\cite{STAR:2025zdq,STAR:2025uxv} with larger acceptance using iTPC detector could directly measure such observables. Furthermore, the observables introduced in this paper should also be particularly useful to be performed as energy dependence of EoS based on the high-baryon density FAIR-CBM, NICA, and HIAF facilities. 

Implementing more self-consistent equations of state with Bayesian analysis methods~\cite{Novak:2013bqa,Bernhard:2019bmu,JETSCAPE:2020mzn} leave as our next work. Recent developments like the Gaussian Process Regression model, which has successfully generated the EoS at zero baryon density~\cite{Gong:2024lhq}, could be particularly valuable as it can be naturally extended to finite baryon densities. While our findings based on the 3D-Glauber initial state and the NEOS-BQS baseline are robust, future work employing different initial condition models and a broader class of finite-$\mu_B$ EoS will also be valuable to further quantify the systematic uncertainties.

\textit{Acknowledgments.}
We thank Jun Xu and Wenhao Zhou for careful reading and valuable comments on the manuscript. This work is supported by the Natural Science Foundation of Shanghai under Grant No. 23JC1400200, the National Key Research and Development Program of China under Contracts Nos. 2022YFA1604900 and 2024YFA1612600, the National Natural Science Foundation of China under Contracts Nos. 12025501, 12225502, and 12147101, Shanghai Pujiang Talents Program under Contract No. 24PJA009, the DOE Research Grant Number DE-SC0024602,
No. DE-SC0021969, DE-SC0024232, and the China Postdoctoral Science Foundation under Grant No. 2024M750489A.
C. S. acknowledge a DOE Office of Science Early Career Award.
The computations in this research are performed at the CFFF platform of Fudan University.

\section*{Supplemental Materials}\label{appendix}
\section{A. Thermodynamic derivation of the speed of sound for the modified EoS} 
Let's start with the pressure in Eq.~(\ref{eq:Palpha}),
\begin{align}
    P_\alpha(T, \mu_B) = P(T, \mu_B) \left(\frac{P}{T^4} \right)^\alpha
    \label{eq:P}
\end{align}
with a tunable parameter $\alpha$. We want to compute the speed of sound from this ansatz and determine the physical range of the parameter $\alpha$ and how $c_s^2$ depends on $\alpha$. 

From pressure, we can compute the other thermodynamic quantities,
\begin{align}
    n_{B, \alpha} &= \frac{\partial P_\alpha}{\partial \mu_B} \bigg\vert_T = (\alpha + 1) \left(\frac{P}{T^4} \right)^\alpha n_B, \label{eq:nB} \\
    s_\alpha &= \frac{\partial P_\alpha}{\partial T} \bigg\vert_{\mu_B} = s \left(\frac{P}{T^4} \right)^\alpha + \alpha \left(\frac{P}{T^4} \right)^\alpha \left(s - 4 \frac{P}{T} \right),  \label{eq:s} \\
    e_\alpha &= T s_\alpha - P_\alpha + n_{B, \alpha} \mu_B = e \left(\frac{P}{T^4} \right)^\alpha + \alpha \left(\frac{P}{T^4} \right)^\alpha (e - 3P) \label{eq:e}.
\end{align}

Eq.~\eqref{eq:nB} shows that the modified net baryon current is proportional to the original one, which means the constraints in NEOS-BQS, $n_Q = 0.4 n_B$ and $n_S = 0$, give the same $\mu_Q$ and $\mu_S$ as functions of $(T, \mu_B)$ as the original one.

If the original equation of state is conformal, $e = 3P$, the modified EoS is conformal as well, $e_\alpha = 3 P_\alpha$. The speed of sound does not change in this limit with $\alpha$.

For EoS with a constant $c_s^2$, $e = \frac{1}{c_s^2} P$, we have
\begin{align}
    e_\alpha &= [e + \alpha (e - 3P)] \left(\frac{P}{T^4} \right)^\alpha \nonumber \\
    &= [e + \alpha (e - 3P)] \frac{P_\alpha}{P} \nonumber \\
    &= \left[\frac{e}{P} + \alpha \left(\frac{e}{P} - 3\right) \right] P_\alpha \nonumber \\
    &= \left[\frac{1}{c_s^2} + \alpha \left(\frac{1}{c_s^2} - 3 \right)\right] P_\alpha.
\end{align}
In this case, the modified speed of sound square is
\begin{align}
    c_{s,\alpha}^2 &= \frac{1}{\left[\frac{1}{c_s^2} + \alpha \left(\frac{1}{c_s^2} - 3 \right)\right]} = \frac{c_s^2}{1 + \alpha (1 - 3 c_s^2)}.
\end{align}
Assuming the original EoS has $0 \le c_s^2 \le 1/3$ and we impose the constraint, $0 \le c_{s,\alpha}^2 \le 1/3$, then we obtain
\begin{align}
    \alpha \ge - 1.
\end{align}
When $\alpha < 0$, $c_{s,\alpha}^2 > c_s^2$ and $\alpha > 0$, $c_{s,\alpha}^2 < c_s^2$.

In the general case, we can compute the speed of sound $c_s^2$,
\begin{align}
    c_s^2 &= \frac{d P}{d e} \bigg\vert_{s/n_B} \nonumber\\
    &= \frac{\frac{\partial P}{\partial T} \left(s \frac{\partial n_B}{\partial \mu_B} - n_B \frac{\partial s}{\partial \mu_B} \right) + \frac{\partial P}{\partial \mu_B} \left(n_B \frac{\partial s}{\partial T} - s \frac{\partial n_B}{\partial T}\right)}{\frac{\partial e}{\partial T} \left(s \frac{\partial n_B}{\partial \mu_B} - n_B \frac{\partial s}{\partial \mu_B} \right) + \frac{\partial e}{\partial \mu_B} \left(n_B \frac{\partial s}{\partial T} - s \frac{\partial n_B}{\partial T}\right)} \nonumber\\
     &= \frac{s \left(s \frac{\partial n_B}{\partial \mu_B} - n_B \frac{\partial s}{\partial \mu_B} \right) + n_B \left(n_B \frac{\partial s}{\partial T} - s \frac{\partial n_B}{\partial T}\right)}{\frac{\partial e}{\partial T} \left(s \frac{\partial n_B}{\partial \mu_B} - n_B \frac{\partial s}{\partial \mu_B} \right) + \frac{\partial e}{\partial \mu_B} \left(n_B \frac{\partial s}{\partial T} - s \frac{\partial n_B}{\partial T}\right)} \nonumber\\
     &= \frac{s^2 \frac{\partial n_B}{\partial \mu_B} - s n_B \left(\frac{\partial s}{\partial \mu_B} + \frac{\partial n_B}{\partial T} \right) + n_B^2 \frac{\partial s}{\partial T}}{\frac{\partial e}{\partial T} \left(s \frac{\partial n_B}{\partial \mu_B} - n_B \frac{\partial s}{\partial \mu_B} \right) + \frac{\partial e}{\partial \mu_B} \left(n_B \frac{\partial s}{\partial T} - s \frac{\partial n_B}{\partial T}\right)}.
    \label{eq:cs2}
\end{align}

From Eq.~\eqref{eq:cs2}, we need to compute the second-order derivatives.

\section{B. Flow decorrelation at $\sqrt{s_{{\mathrm{NN}}}} = 19.6~\text{GeV}$}
\begin{figure}[ht]
\centering
\vspace*{-0.2cm}
\includegraphics[width=0.95\linewidth]{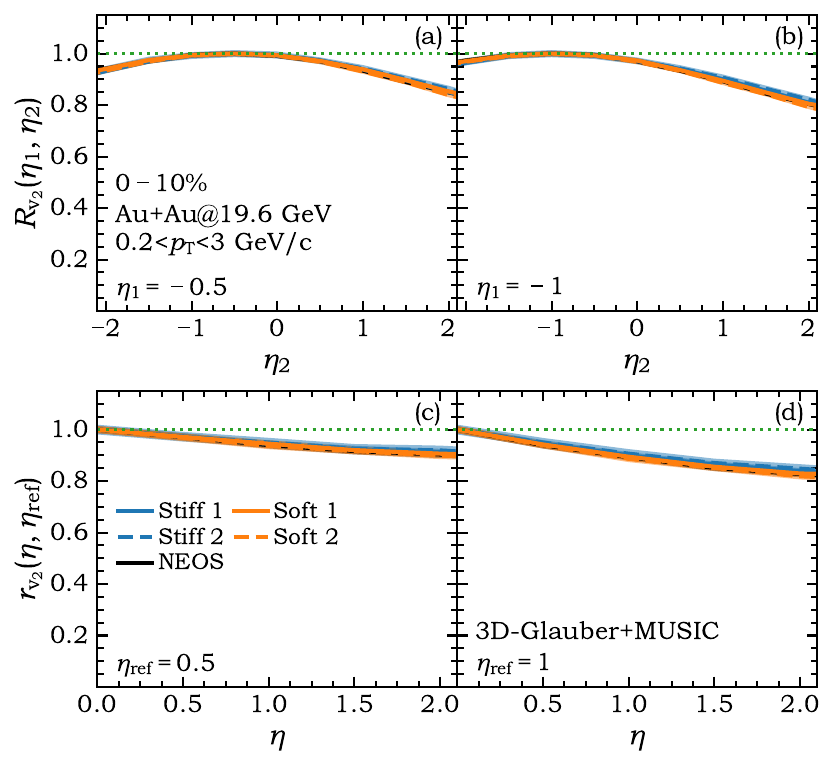}
\vspace*{-0.5cm}
\caption{$v_2$ decorrelation coefficients $R_{v_2}$ and $r_{{v_2}}$ in 0--10\% central $^{197}$Au+$^{197}$Au collisions at $\sqrt{s_{\rm NN}}=$ 19.6 GeV. Panels (a) and (b) show $R_{v_2}(\eta_1, \eta_2)$ with $\eta_1 = -0.5$ and $\eta_1 = -1$, respectively; Panels (c) and (d) present $r_{v_2}(\eta, \etaref)$ with $\etaref = 0.5$ and $\etaref = 1$, respectively.}\label{fig:Rv2}
\end{figure}
Fig.~\ref{fig:Rv2} presents the $v_2$ decorrelation in 0--10\% central $^{197}$Au+$^{197}$Au collisions at $\sqrt{s_{_{\mathrm{NN}}}} = 19.6$ GeV. Unlike the $p_T$ decorrelation, which exhibits a deviation from unity of approximately 50\% at $\Delta\eta \sim 2$ as shown in Fig.~\ref{fig:capRpT}, the flow decorrelation is significantly weaker and remains almost unaffected by the EoS and viscosity.
In event-by-event collisions,  anisotropic flow exhibit fluctuations along the rapidity direction within a single collision event~\cite{Jia:2014ysa,Pang:2015zrq}. 
The resulting decorrelation can be quantified as~\cite{Zhang:2024bcb,Jia:2024xvl}
\begin{eqnarray}\label{eq:Rvn}
R_{v_n}(\eta_{1}, \eta_{2}) &=& \frac{\langle {V}_{n}(\eta_{1})  {V}_{n}^*(\eta_{2}) \rangle}{\sqrt{\langle {V}_{n}(\eta_{1})  {V}_{n}^*(\eta_{1}) \rangle\langle {V}_{n}(\eta_{2})  {V}_{n}^*(\eta_{2}) \rangle}} \\
r_{{v_n}}\left({\eta},{\etaref}\right)&=&\frac{\langle {V}_{n}(-\eta)  {V}_{n}^*(\etaref) \rangle}{\langle {V}_{n}(\eta)  {V}_{n}^*(\etaref) \rangle} = \frac{R_{{v_n}}\left({-\eta},{\etaref}\right)}{R_{{v_n}}\left({\eta},{\etaref}\right)}
\end{eqnarray}
with \( V_{n} = v_{n} e^{in\Psi} \) is vector form of the event-by-event anisotropic flow. 
In the latter definition, $r_{v_n}$ is often assumed to be independent of the reference pseudorapidity $\eta_\text{ref}$; however, this assumption has been shown to be inadequate using transport model calculations in Ref.~\cite{Zhang:2024bcb}.

\section{C. $p_T$ decorrelation at $\sqrt{s_{{\mathrm{NN}}}} = 200~\text{GeV}$}
\begin{figure}[ht]
\centering
\vspace*{-0.2cm}
\includegraphics[width=0.95\linewidth]{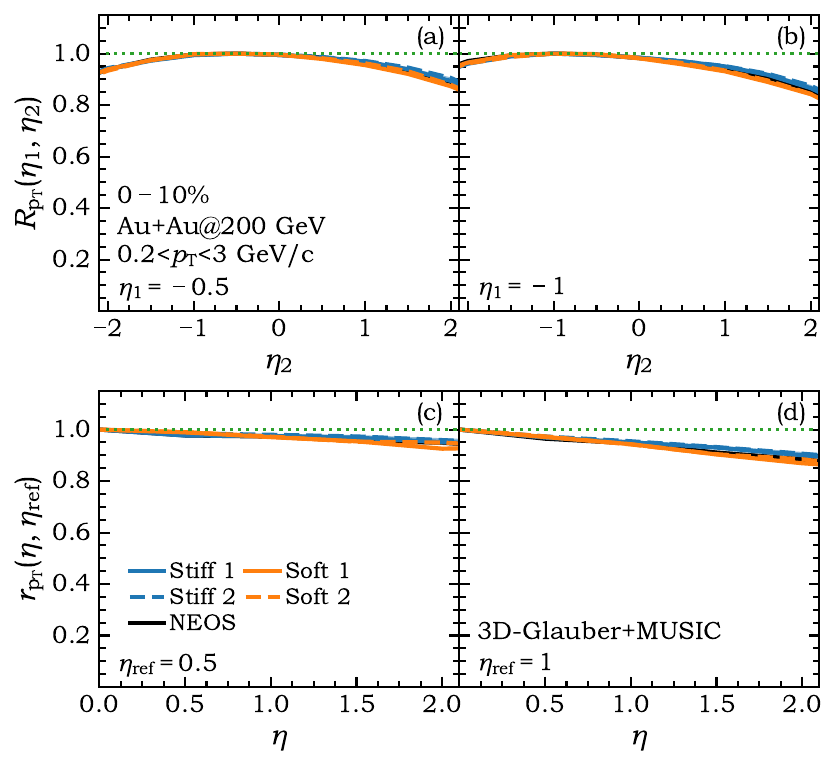}
\vspace*{-0.5cm}
\caption{Mean transverse momentum decorrelation coefficients in 0--10\% central $^{197}$Au+$^{197}$Au collisions at $\sqrt{s_{\mathrm{NN}}} = 200$ GeV. Panels (a) and (b) show $R_{p_T}(\eta_1, \eta_2)$ with $\eta_1 = -0.5$ and $\eta_1 = -1$, respectively; Panels (c) and (d) present $r_{p_T}(\eta, \etaref)$ with $\etaref = 0.5$ and $\etaref = 1$, respectively.}\label{fig:RpT_200}
\end{figure}

We have also computed the $p_T$ decorrelation in 0-10\% central $^{197}$Au+$^{197}$Au collisions at $\sqrt{s_{\mathrm{NN}}} = 200$ GeV, as shown in Fig.~\ref{fig:RpT_200}. The results indicate that the EoS effect is difficult to observe up to a pseudorapidity gap of $\Delta \eta = 2$, which can be attributed to the large beam rapidity at this collision energy.

\bibliography{ref}
\end{document}